\documentclass[preprint,12pt,3p]{elsarticle} 

\usepackage{pifont}
\usepackage{natbib}
\usepackage{geometry}
\usepackage{fleqn}
\usepackage{txfonts}
\usepackage{hyperref}

\usepackage{amssymb}
\usepackage{graphicx}
\usepackage{amsbsy}
\usepackage{color}

\newcommand{\bq}{\begin{eqnarray}}
\newcommand{\eq}{\end{eqnarray}}
\newcommand{\bqn}{\begin{eqnarray*}}
\newcommand{\eqn}{\end{eqnarray*}}
\newcommand{\SSS}{{\bf S}}

\newcommand{\rr}{{\bf r}}
\newcommand{\RR}{{\bf R}}

\newcommand{\nablab}{\pmb{\nabla}}

\begin{document}
\title{Hellmann and Feynman theorem versus diffusion Monte Carlo experiment}

\author[ric]{Riccardo Fantoni\corref{corr}}
\ead{rfantoni27@sun.ac.za}
\ead[url]{http://www-dft.ts.infn.it/~rfantoni/}
\address[ric]{National Institute for Theoretical Physics (NITheP) and
Institute of Theoretical Physics, Stellenbosch 7600, South Africa} 
\cortext[corr]{Corresponding author}

\date{\today}

\begin{abstract}
In a computer experiment the choice of suitable estimators to measure
a physical quantity plays an important role. We propose a new direct
route to determine estimators for observables which do not commute
with the Hamiltonian. Our new route makes use of the Hellmann and
Feynman theorem and in a diffusion Monte Carlo simulation it
introduces a new bias to the measure due to the choice of the
auxiliary function. This bias is independent from the usual one due to 
the choice of the trial wave function. We used our route to measure
the radial distribution function of a spin one half Fermion fluid.
\end{abstract}


\maketitle
An important component of a computer experiment of a many particles
system, a fluid, is the determination of suitable {\sl estimators} 
to measure, through a statistical average, a 
given physical quantity, an observable. Whereas the average from
different estimators must give the same result, the variance, the
square of the statistical error, can be different for different
estimators. We will denote with $\langle {\cal O}\rangle_f$ the
measure of the physical observable ${\cal O}$ and with
$\langle\ldots\rangle_f$ the statistical average over the  
probability distribution $f$. In this communication we use the word
estimator to indicate the function ${\cal O}$ itself, unlike the
more common use of the word to indicate the usual Monte Carlo
estimator $\sum_{i=1}^{\cal N}{\cal O}_i/{\cal N}$ of the average,
where $\{{\cal O}_i\}$ is the set obtained evaluating ${\cal O}$ over a
finite number ${\cal N}$ of points distributed according to $f$.
This aspect of finding out different ways of calculating quantum
properties in some ways resembles experimental physics. The
theoretical concept may be perfectly well defined but it is up to the
ingenuity of the experimentalist to find the best way of doing the
measurement. Even what is meant by ``best'' is subject to debate.

In ground state Monte Carlo simulations \cite{McMillan1965,Kalos1974},
unlike classical Monte Carlo 
simulations \cite{Hockney,Allen-Tildesley,Frenkel-Smit} and path
integral Monte Carlo simulations \cite{Ceperley1995}, one has to 
resort to the use of a trial wave function 
\cite{McMillan1965}, $\Psi$. While this is not a source of error, {\sl
  bias}, in a
diffusion Monte Carlo simulation \cite{Kalos1974} of a system of
Bosons, it is for a system of Fermions, due to the {\sl sign problem}
\cite{Ceperley1991}. Since this is always present in a Monte Carlo
simulation of Fermions we will not consider any further when
talking about the bias.   

Another source of bias inevitably present in all three experiments,
which we will not take into consideration in the following, is
the finite size error. In the rest of the paper we will generally
refer to the bias to indicate the error (neglecting the finite size
error and the sign problem) that we make when defining different
estimators of the same quantity not giving the same average.

In a ground state Monte Carlo simulation, the energy has the
{\sl zero-variance} principle \cite{Ceperley1979}: as the trial wave
function approaches the exact ground state, the statistical error
vanishes. In a diffusion Monte Carlo simulation of a system of Bosons
the local energy of the trial wave function,
$E_L(\RR)=[H\Psi(\RR)]/\Psi(\RR)$, where $\RR$ denotes a configuration
of the system of particles and $H$ is the Hamiltonian assumed to be
real, is an unbiased estimator for the ground state. For 
Fermions, the ground state energy measurement is biased by the sign
problem. For observables $O$ which do not commute with the Hamiltonian,
the local estimator, $O_L(\RR)=[O\Psi(\RR)]/\Psi(\RR)$, is inevitably
biased by the choice of the 
trial wave function. A way to remedy to this bias can be the use of
the forward walking method \cite{Kalos1974b,Barnett1991} or the
reptation quantum Monte Carlo method \cite{Baroni1999} to reach pure
estimates. Otherwise this bias can be made of leading order
$\delta^2$, with $\delta=\phi_0-\Psi$ where $\phi_0$ is the ground
state wave function, introducing the extrapolated measure,
$\overline{O}^{\rm ext}=2\langle O_L\rangle_f-\langle
O_L\rangle_{f_{\rm vmc}}$
where the first statistical average, the {\sl mixed measure}, 
is over the diffusion Monte Carlo (DMC) stationary
probability distribution $f$ and the second,
the {\sl variational measure}, over the variational Monte
Carlo (VMC) probability distribution $f_{\rm vmc}$ which can also be
obtained as the stationary probability distribution of a DMC without
branching \cite{Umrigar1993}.

One may follow different routes to determine estimators such as the {\sl
  direct} microscopic route, the {\sl virial} route through the use of
the virial theorem, or the {\sl thermodynamic} route through the use
of thermodynamic identities. In an unbiased experiment the different
routes to the same observable must give the same average. 

In this communication we propose to use the Hellmann and Feynman
theorem as a direct route for the determination of estimators in a
diffusion Monte Carlo simulation. Some attempts in this direction have
been tried before \cite{Assaraf2003,Gaudoin2007}. The novelty of our
approach, respect to Ref. \cite{Assaraf2003}, is a different definition
of the correction to the variational measure, necessary in the
diffusion experiment, and, respect to Ref. \cite{Gaudoin2007}, the
fact that the bias stemming from the sign problem does not exhaust all
the bias due to the choice of the trial wave function.  

We start with the
eigenvalue expression $(H^\lambda-E^\lambda)\Psi^\lambda=0$ for the
ground state of the perturbed Hamiltonian $H^\lambda=H+\lambda O$,
take the derivative with respect to the parameter $\lambda$, multiply
on the right by the ground state at $\lambda=0$, $\phi_0$, and
integrate over the particles configuration to get 
\bq \nonumber
\int d\RR\,\phi_0(H^\lambda-E^\lambda)\frac{\partial\Psi^\lambda}
{\partial\lambda}=
\int d\RR\,\phi_0
\left(\frac{d E^\lambda}{d\lambda}-
\frac{d H^\lambda}{d\lambda}\right)\Psi^\lambda~.
\eq 
Then we note that due to the Hermiticity of the Hamiltonian the left
hand side vanishes at $\lambda=0$ so that we get further
\bq
\left.\frac{\int d\RR\,\phi_0 O\Psi^\lambda}{\int
  d\RR\,\phi_0\Psi^\lambda}\right|_{\lambda=0}=
\left.\frac{d E^\lambda}{d\lambda}\right|_{\lambda=0}~.
\eq
This relation holds only in the $\lambda\to 0$ limit
unlike the more common form \cite{LandauQM} which holds
for any $\lambda$. Given $E^\lambda=\int
d\RR\phi_0H^\lambda\Psi^\lambda/\int d\RR\phi_0\Psi^\lambda$ the
``Hellmann and Feynman'' (HF) measure in a diffusion Monte Carlo
experiment is then defined as follows 
\bq \label{zvzb-d}
\overline{O}^{\rm HF}&=&\left.\frac{dE^\lambda}{d\lambda}\right|_{\lambda=0}
\approx \langle O_L(\RR)\rangle_{f}+\langle\Delta
O_L^{\alpha}(\RR)\rangle_{f}+\langle\Delta O_L^{\beta}(\RR)\rangle_{f}~.
\eq 
The $\alpha$ correction is  
\bq \label{zv-d}
\Delta
O_L^{\alpha}(\RR)=\left[\frac{H\Psi^\prime(\RR)}{\Psi^\prime(\RR)}
-E_L(\RR)\right]\frac{\Psi^\prime(\RR)}{\Psi(\RR)}~.
\eq
In a variational Monte Carlo experiment
this term, usually, does not contribute to the average (with respect to
$f_{\rm vmc}\propto \Psi^2$) due to the Hermiticity of the Hamiltonian. 
We will then define a Hellmann and Feynman variational (HFv)
estimator as $O^{\rm HFv}=O_L+\Delta O_L^\alpha$. The $\beta$
correction is  
\bq \label{zb-d}
\Delta O_L^{\beta}(\RR)=[E_L(\RR)-E_0]\frac{\Psi^\prime(\RR)}{\Psi(\RR)}~,
\eq
where $E_0=E^{\lambda=0}$ is the ground state energy. It should be
noticed that our correction differs by a factor $1/2$ from the
zero-bias correction defined  in Ref. \cite{Assaraf2003} because these
authors chose $E^\lambda=\int
d\RR\Psi^\lambda H^\lambda\Psi^\lambda/\int
d\RR(\Psi^\lambda)^2$ right from the start. This correction is 
necessary in a diffusion Monte Carlo experiment not to bias the
measure. The extrapolated Hellmann and Feynman measure will then be
$\overline{O}^{\rm HF-ext}=2\overline{O}^{\rm HF}-\langle
O^{\rm HFv}\rangle_{f_{\rm vmc}}$. Both corrections $\alpha$ and
$\beta$ to the local estimator depend on the {\sl auxiliary} function, 
$\Psi^\prime=\partial\Psi^\lambda/\partial\lambda|_{\lambda=0}$.
Of course if, on the left hand side of Eq. (\ref{zvzb-d}), we had
chosen $\Psi^{\lambda=0}$ as the exact ground state wave function,
$\phi_0$, instead of the trial wave function, $\Psi$, then both
corrections would have 
vanished. When the trial wave function is sufficiently close to the
exact ground state function a good approximation to the auxiliary
function can be obtained from first order perturbation theory for
$\lambda\ll 1$. So the Hellmann and Feynman measure is affected by the
new source of bias due to the choice of the auxiliary function which is
independent from the bias due to the choice of the trial wave function.  

We applied the Hellmann and Feynman route to the measurement of the
radial distribution function (RDF) of the Fermion fluid
studied by Paziani \cite{Paziani2006}. This is a fluid of spin
one-half particles interacting with a bare pair-potential
$v_\mu(r)={\rm erf}(\mu r)/r$ immersed in a ``neutralizing''
background. The pair-potential depends on the parameter $\mu$ in such
way that in the limit $\mu\to 0$ one recovers the ideal Fermi gas and
in the limit $\mu\to\infty$ one finds the Jellium model. We chose this
model because it allows to move continuously from a situation where
the trial wave 
function coincides with the exact ground state, in the $\mu\to 0$
limit, to a situation where the correlations due to the particles
interaction become important, in the opposite $\mu\to \infty$ limit.

We chose as auxiliary function $\Psi^\prime=Q\Psi$, the first one of
Toulouse {\sl et al.} \cite{Toulouse2007} (their Eq. (30)),
\bq \label{Toulouse}
Q_{\sigma,\sigma^\prime}(r,\RR)=
-\frac{r_s^2}{8\pi V
n_{\sigma}n_{\sigma^\prime}}\sum_{i,j\neq i}\delta_{\sigma,\sigma_i}
\delta_{\sigma^\prime,\sigma_j}\int\frac{d\Omega_\rr}{4\pi}
\frac{1}{|\rr-\rr_{ij}|}~,
\eq
here $\sigma$ and $\sigma^\prime$ denote the spin species, $r=|\rr|$
the separation between two particles, $r_{ij}$ the separation between
particle $i$ and $j$, $\sigma_i$ the spin species of particle $i$,
and $d\Omega_\rr$ is the solid angle element of integration. The
particles are in a recipient of volume $V$ at a density
$n=n_++n_-=1/[4\pi(a_0 r_s)^3/3]$ with $a_0$ the Bohr radius,
$a=a_0r_s$ the lengths unit, and $n_\sigma$ the density of the spin
$\sigma$ particles.  
With this choice the $\alpha$ correction partially cancels the
histogram estimator 
$I_{\sigma,\sigma^\prime}(r,\RR)=\sum_{i,j\neq
  i}\delta_{\sigma,\sigma_i} 
\delta_{\sigma^\prime,\sigma_j}\int
\delta(\rr-\rr_{ij})\,d\Omega_\rr/(4\pi V
n_{\sigma}n_{\sigma^\prime})$,   
and one is left with a HFv estimator which goes to zero at large
$r$. This is because the quantity
$\langle\Delta I^{\alpha}_{\sigma,\sigma^\prime}(r,\RR)\rangle_{\Psi^2}$
$=-\int_{\partial V^N}\Psi^2(\RR)\nablab_\RR
Q_{\sigma,\sigma^\prime}(r,\RR)\cdot d\SSS/r_s^2$ equals minus one for
all $r$ with $\rr\in V$, instead of zero 
as normally expected. This is ultimately related to the behavior of
the auxiliary function on the border of $V^N$. 
The measure of the $\beta$ correction also goes to zero at
large $r$ because one is left with a statistical average of a quantity
proportional to $E_L(\RR)-E_0$. The Hellmann and Feynman measure
needs then to be shifted by $+1$.

Our variational Monte Carlo experiments showed that in the variational
measure the average of the 
histogram estimator agrees with the average of the HFv estimator
within the square root of the variance of the average
$\sigma_{\rm av}=\sqrt{\sigma^2{\cal K}/{\cal N}}$ (here $\sigma^2$
is the variance, ${\cal K}$ the correlation time of the random walk,
and ${\cal N}$ the number of Monte Carlo steps) and the two
$\sigma_{\rm av}$ are comparable. This is expected since the HFv
estimator is defined exactly as in Ref. \cite{Assaraf2003} which
correctly takes into account the definition of the HF estimator within
a variational Monte Carlo simulation. In the fixed nodes diffusion
experiment, where one has to add the $\beta$ correction 
not to bias the average (note once again that this is defined by us as
one half the zero-bias correction of Ref. \cite{Assaraf2003}), the
Hellmann and Feynman measure has an 
average in agreement with the one of the histogram estimator but the
$\sigma_{\rm av}$ increases. This is to be expected from the
extensive nature of the 
$\beta$ correction in which the energy appears. Of course the averages
from the extrapolated Hellmann and Feynman measure and the extrapolated
measure for the histogram estimator also agree. 

In the simulation for the Coulomb case, $\mu\to\infty$, we made
extrapolations in time step and number of walkers for each value of
$r_s$. Given a relative precision $\delta_{e_0}=\Delta e_0/e_p^x $, where 
$e_0=\langle E_L\rangle_{f}/N$, $\Delta e_0$ is the statistical error on $e_0$,
and $e_p^x$ is the exchange energy, we set as our target relative
precision $\delta_{e_0}=10^{-2}\%$. The extrapolated values of the
time step and number of walkers were then used for all other values of
$\mu$. We chose the trial wave function of the Bijl-Dingle-Jastrow
\cite{Bijl40,Dingle49,Jastrow55} form as a product of Slater
determinants and a Jastrow factor. The pseudo-potential was chosen as in
Ref. \cite{Ceperley2004}, ${\cal J}_2$, 
which is expected to give better results for Jellium. Comparison with
the simulation of the unpolarized fluid at $r_s=1$ and $\mu=1$ with
the pseudo potential of Ref. \cite{Ceperley78}, ${\cal J}_1$, for
which the trial wave 
function becomes the exact ground state wave function in the $\mu\to
0$ limit, shows that the two extrapolated measures of the unlike
histogram estimator differ one from the other by less than $7\times 10^{-3}$,
the largest difference being at contact (see the inset of Fig. 1). The
use of more 
sophisticated trial wave functions, taking into account the effect 
of backflow and three-body correlations, is found to affect the
measure by even less. In Table \ref{tab:got_z0} we compare the contact
values of the unlike RDF of the unpolarized fluid at various $r_s$ and
$\mu$ from the measures of the histogram estimator and the HF
measures. We see that there is disagreement between the measure from
the histogram estimator and the HF measure only in the Coulomb
$\mu\to\infty$ case at $r_s=1,2$.

\begin{figure}[H]
\begin{center}
\includegraphics[width=8cm]{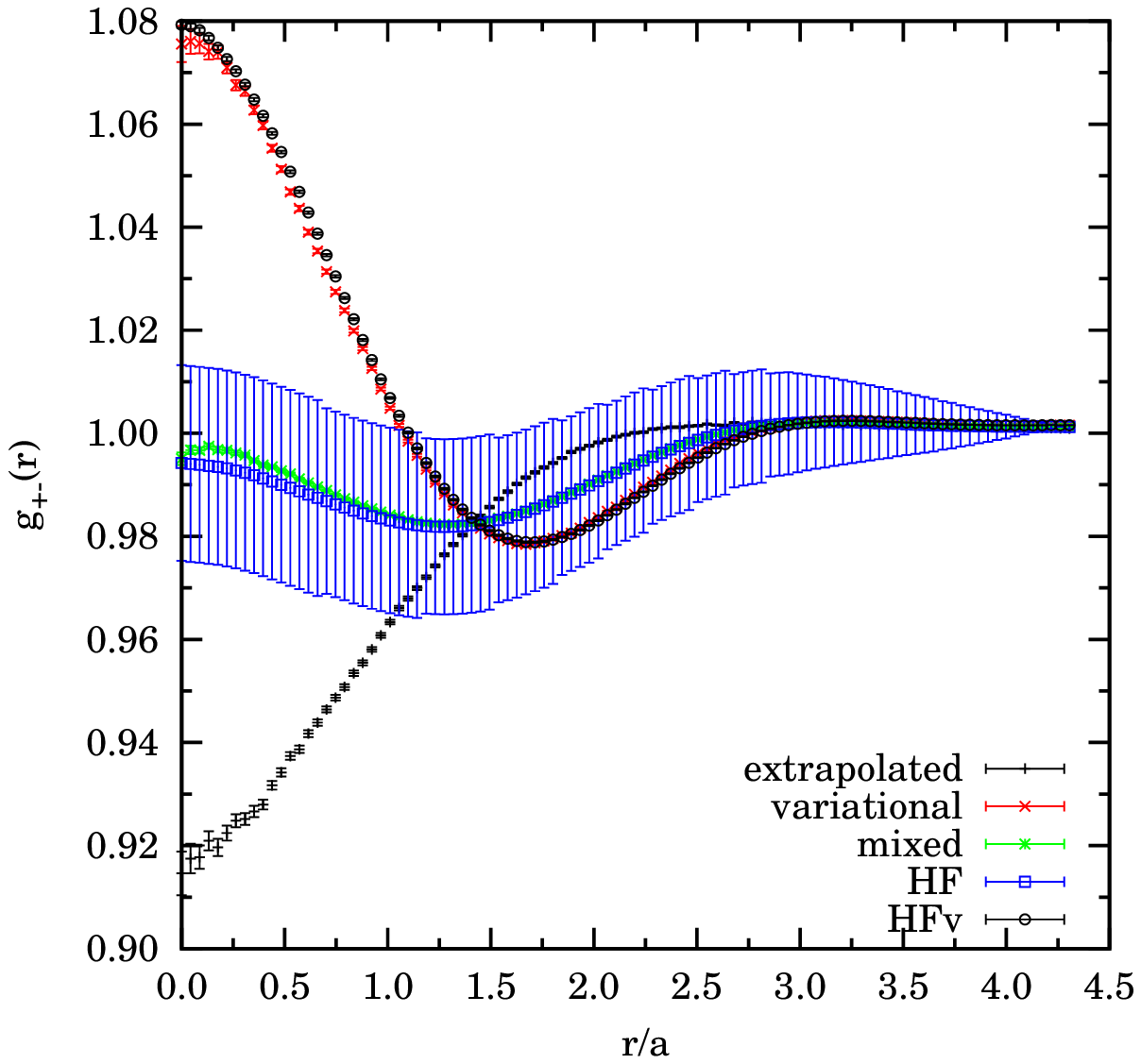}
\includegraphics[width=8cm]{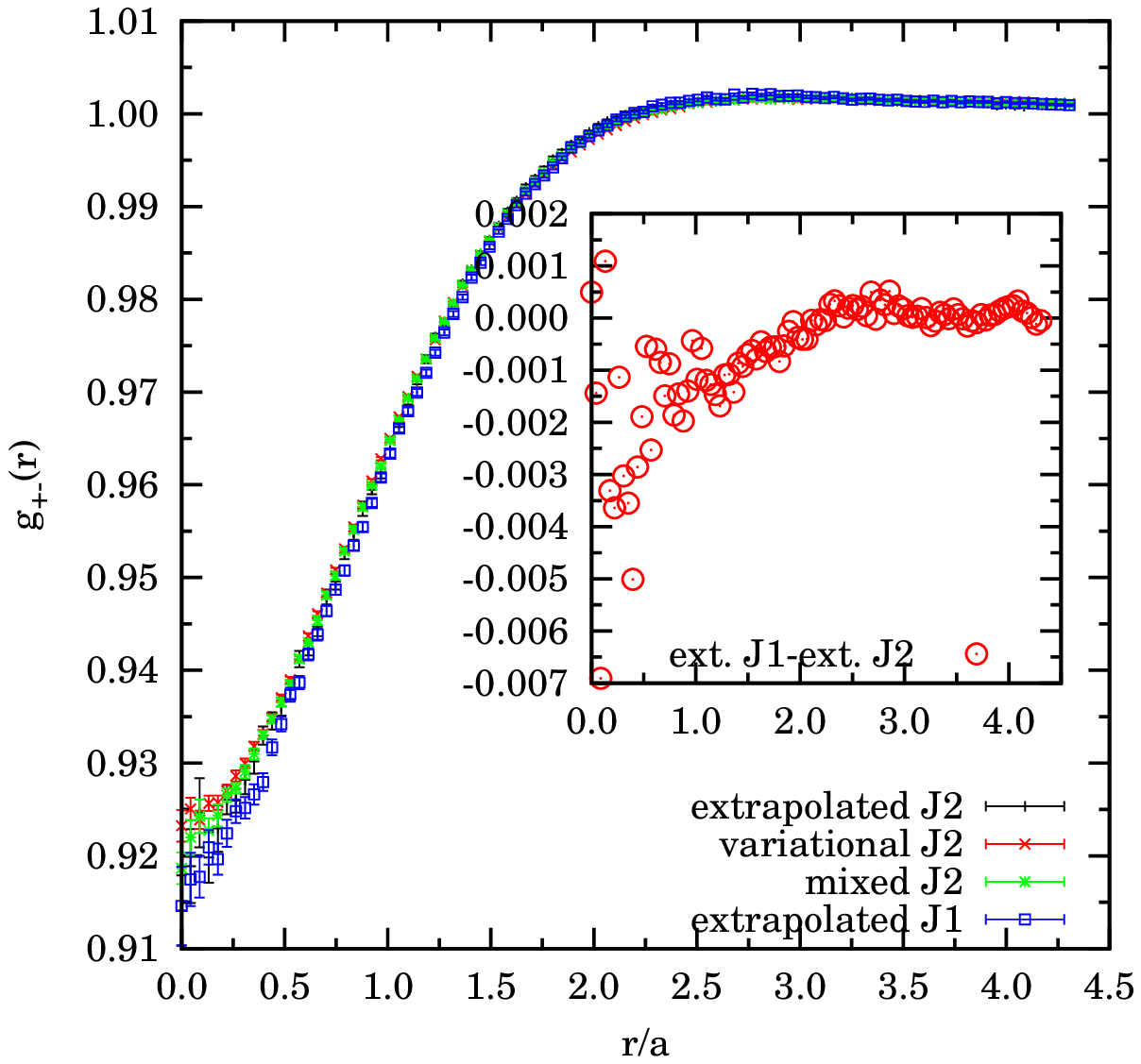}
\end{center}
\caption{Unlike RDF for the unpolarized fluid of Paziani
  \cite{Paziani2006} at $r_s=1$ and $\mu=1$ with $162$ particles. On
  the left panel the calculation with the Jastrow ${\cal J}_1$ with
  various measures: 
  variational histogram (variational) and variational HFv (HFv), mixed
  histogram (mixed) and HF (HF), and 
  extrapolated histogram (extrapolated). On the right panel the
  calculation with the Jastrow ${\cal J}_2$ with the histogram
  variational (variational ${\cal J}_2$), 
  mixed (mixed ${\cal J}_2$), and extrapolated (extrapolated ${\cal J}_2$)
  measures. Also the extrapolated measure with the Jastrow ${\cal
    J}_2$ is compared with the extrapolated measure with the Jastrow
  ${\cal J}_1$. In the
  inset is shown the difference between the histogram extrapolated
  measure of the calculation with ${\cal J}_1$ and the histogram extrapolated
  measure of the calculation with ${\cal J}_2$. $10^5$ Monte Carlo steps were
  used in the simulations.} 
\label{fig:cusp}
\end{figure}
\begin{table}
\caption{Contact values for the unlike RDF of the unpolarized fluid of
  Paziani \cite{Paziani2006}
  at various $r_s$ and $\mu$ from the mixed
  measure of the histogram estimator (hist) and the HF
  measure (HF) with the auxiliary function chosen as in
  Eq. (\ref{Toulouse}), also reported are the two 
  extrapolated measures (ext and HF-ext). The trial wave function used
  was of the Slater-Jastrow type with the Jastrow of
  Ref. \cite{Ceperley2004}, ${\cal J}_2$. 
  The last column gives the error on the HF
  measure. $162$ particles were used with $10^5$ Monte Carlo steps.
\label{tab:got_z0}}
{\scriptsize
\begin{tabular} {|ccccccc|}
\hline
$r_s$& $\mu$ & hist & ext & HF & HF-ext & $\sigma_{\rm av}$ on HF\\
\hline
10&1/2&1.000(4)&0.91(1)&1.00&0.92&0.03\\
10&1&0.644(3)&0.582(8)&0.65&0.59&0.03\\
10&2&0.182(1)&0.146(4)&0.18&0.14&0.06\\
10&4&0.0506(8)&0.048(2)&0.05&0.04&0.07\\
10&$\infty$&0.0096(3)&0.0118(8)&0.00&0.00&0.09\\
\hline
5&1/2&1.034(3)&0.94(1)&1.03&0.94&0.03\\
5&1&0.796(3)&0.743(9)&0.79&0.73&0.02\\
5&2&0.405(2)&0.362(6)&0.40&0.36&0.02\\
5&4&0.199(1)&0.184(4)&0.20&0.18&0.03\\
5&$\infty$&0.0799(8)&0.080(2)&0.06&0.06&0.03\\
\hline
2&1/2&1.0618(4)&0.97(1)&1.05&0.95&0.04\\
2&1&0.927(3)&0.852(9)&0.93&0.86&0.03\\
2&2&0.697(3)&0.639(9)&0.69&0.63&0.02\\
2&4&0.511(2)&0.473(7)&0.51&0.47&0.02\\
2&$\infty$&0.349(2)&0.323(5)&0.32&0.30&0.02\\
\hline
1&1/2&1.077(3)&0.98(1)&1.07&0.97&0.02\\
1&1&0.994(3)&0.91(1)&0.99&0.91&0.02\\
1&2&0.855(3)&0.787(9)&0.86&0.81&0.02\\
1&4&0.730(2)&0.676(8)&0.73&0.66&0.01\\
1&$\infty$&0.602(2)&0.560(7)&0.58&0.53&0.01\\
\hline
\end{tabular}
}
\end{table}

In conclusions we defined a Hellmann and Feynman estimator to measure
a given physical property either in a variational Monte Carlo
experiment and in a diffusion Monte Carlo experiment. Our definition
coincides with the one of Ref. \cite{Assaraf2003} in the variational
case but is different in the diffusion case. We proof tested our
definitions on the calculation of the radial distribution function of
a particular Fermion fluid. Our simulations showed that the bias is
correctly accounted for in both kind of experiments but the variance
increases in the diffusion experiment relative to the one of the
histogram estimator.
We believe it is still an open problem the one of determining the
relationship between the choice of the auxiliary function and the
variance of the Hellmann and Feynman measure.

The idea for the work came from discussions with Saverio Moroni. I
would also like to acknowledge the hospitality of the National
Institute for Theoretical Physics (NITheP) of South Africa where the
work was done. 
The simulations were carried out at the Center for High Performance
Computing (CHPC), CSIR Campus, 15 Lower Hope St., Rosebank, Cape Town,
South Africa. 

\end{document}